\documentclass{article}

\usepackage{graphicx,natbib,amssymb,lineno,hyperref}
\usepackage{lscape}

\def\astrobj#1{#1}

\newcommand\apj{{ApJ}}%
\newcommand\aj{{AJ}}%

\begin{document}

\begin{center}
 \LARGE{\textbf{The data mining II: An analysis of 33 eclipsing binary light-curves observed by the INTEGRAL/OMC}} \\[5mm]
 {P. Zasche} \\[5mm]
 \Large{Instituto de Astronom\'{\i}a, Universidad Nacional Aut\'onoma de M\'exico, A.P. 70-264, M\'exico,
DF 04510, Mexico}\\[4mm]

{Astronomical Institute, Faculty of Mathematics and Physics, Charles University Prague, CZ-180 00
Praha 8, V Hole\v{s}ovi\v{c}k\'ach 2, Czech Republic}\\
{email: zasche@sirrah.troja.mff.cuni.cz}\\[4mm]

\end{center}

{\normalsize \textbf{Abstract:} Thirty-three eclipsing binaries were selected for an analysis from
a huge database of observations made by the INTEGRAL/OMC camera. The photometric data were
processed and analyzed, resulting in a first light-curve study of these neglected eclipsing
binaries. The system CY~Lac was discovered to be an eccentric one. In several systems from this
sample even their orbital periods have been confirmed or modified. Due to missing spectroscopic
study of these stars, further detailed analyses are still needed.}\\

\astrobj{V964 Aql}, \astrobj{V1426 Aql}, \astrobj{V1450 Aql}, \astrobj{LP Ara}, \astrobj{DQ Car},
\astrobj{DR Car}, \astrobj{BZ Cas}, \astrobj{V654 Cas}, \astrobj{PQ Cen}, \astrobj{V379 Cen},
\astrobj{V Cir}, \astrobj{DO Cyg}, \astrobj{DP Cyg}, \astrobj{V536 Cyg}, \astrobj{V537 Cyg},
\astrobj{V616 Cyg}, \astrobj{V642 Cyg}, \astrobj{V703 Cyg}, \astrobj{V359 Her}, \astrobj{CY Lac},
\astrobj{YY Nor}, \astrobj{HM Nor}, \astrobj{V537 Oph}, \astrobj{BS Sco}, \astrobj{V569 Sco},
\astrobj{V714 Sco}, \astrobj{BN Sgr}, \astrobj{V780 Sgr}, \astrobj{V2168 Sgr}, \astrobj{XY Vel},
\astrobj{YY Vel}, \astrobj{AZ Vel} ; stars: fundamental parameters ; PACS codes: 97.10.-q ;
97.80.-d ; 97.80.Hn

\section{Introduction}

The INTEGRAL (INTErnational Gamma-Ray Astrophysics Laboratory) satellite produces many observations
since its launch in 2002, not only in gamma part of the spectra. The onboard OMC (Optical
Monitoring Camera) was designed to obtain the observations in optical $V$ passband. These
observations are in fact only a by-product of the mission, but nowadays there are many observations
available.

Despite the fact that the database of these measurements is freely available on internet, the
analyses are still very rare. The most recent one using the OMC data is that by
\cite{2009IBVS.5881....1J} about a new $\beta$~Cep star.

This investigation is directly following our previous papers (\citealt{Zasche2008NewA} and
\citealt{2009NewA...14..129Z}). The selection criteria used here were also the same: maximum number
of data points and non-existence of any detailed light-curve analysis of the particular system.
There were 33 systems selected for the present paper.

\section{Analysis of the individual systems}

All observations of these systems were carried out by the same instrument (50mm OMC telescope) and
the same filter (standard Johnson's V filter). Time span of the observations ranges from November
2002 to October 2008. A transformation of the time scale has been done following the equation
$Julian Date - ISDC Julian Date = 2,451,544.5$. Only a few outliers from each data set were
excluded. The {\sc Phoebe} programme (see e.g. \citealt{Prsa2005}), based on the Wilson-Devinney
algorithm \citep{Wilson1971}, was used for the analysis.

Due to missing information about the stars, and having only the light curves in one filter, some of
the parameters have to be fixed. At first, for all systems we have used the "Detached binary" mode
(in Wilson \& Devinney mode 2) and also the "Semidetached with the secondary component filling its
Roche lobe" (mode 5 in Wilson \& Devinney) for computing. For both modes a "q-search method" was
used, which means trying to find the best fit with different values of the mass ratio $q$ ranging
from 0 to 1 with a step 0.1. The limb-darkening coefficients were interpolated from van~Hamme's
tables (see \citealt{vanHamme1993}), the linear cosine law was used. The values of the gravity
brightening and bolometric albedo coefficients were set at their suggested values for convective
atmospheres (see \citealt{Lucy1968}), i.e. $G_1 = G_2 = 0.32$, $A_1 = A_2 = 0.5$. In all cases
(except for CY~Lac) the orbital eccentricity was set to 0 (circular orbit). Therefore, the
quantities which could be directly calculated from the light curve are the following: The
luminosity ratio $L_1/L_2$, the temperature ratio $T_1/T_2$, the inclination $i$, ephemerides of
the system, the Kopal's modified potentials $\Omega_1$ and $\Omega_2$, the synchronicity parameters
$F_1$ and $F_2$, the third light $l_3$, and the mass ratio $q$. Using the parameters introduced
above, one could also derive the value of the radii ratio $R_1/R_2$.

The distinguishing between the minima has been done only according to the observational point of
view, which means that the deeper one is the primary one. This results in a fact that the primary
component could be neither the larger one, nor the more massive one. In two cases the secondary
components result to be the more luminous ones (V1450~Aql and V714 Sco), and in several cases also
the more massive ones.

All of the basic information about the analyzed systems are introduced in Table \ref{Table1}, where
are the $B$ and $V$ magnitudes from the GCVS (\citealt{1971GCVS} and \citealt{Malkov}), the $B-V$
values from the GCVS and also from the NOMAD catalogue \citep{NOMAD2004}. The spectral types are
taken from the published literature and also from the \citeauthor{Svechnikov} (S\&K,
\citealt{Svechnikov}). The estimated mass ratio and also the type of the eclipsing binary have been
taken from S\&K (EA stands for the Algol type, while EB for the $\beta$~Lyrae type, SD for
semi-detached systems, DS for detached ones with subgiant secondary, and DM for detached main
sequence ones). 'Minima' stands for the number of published times of minima and the last four
columns introduce the actual OMC magnitudes in Johnson's $V$ filter, the depths of both primary and
also secondary minima in $V$ filter, and finally the number of data points used for this analysis.

\begin{figure}[b]
 \includegraphics[width=16cm]{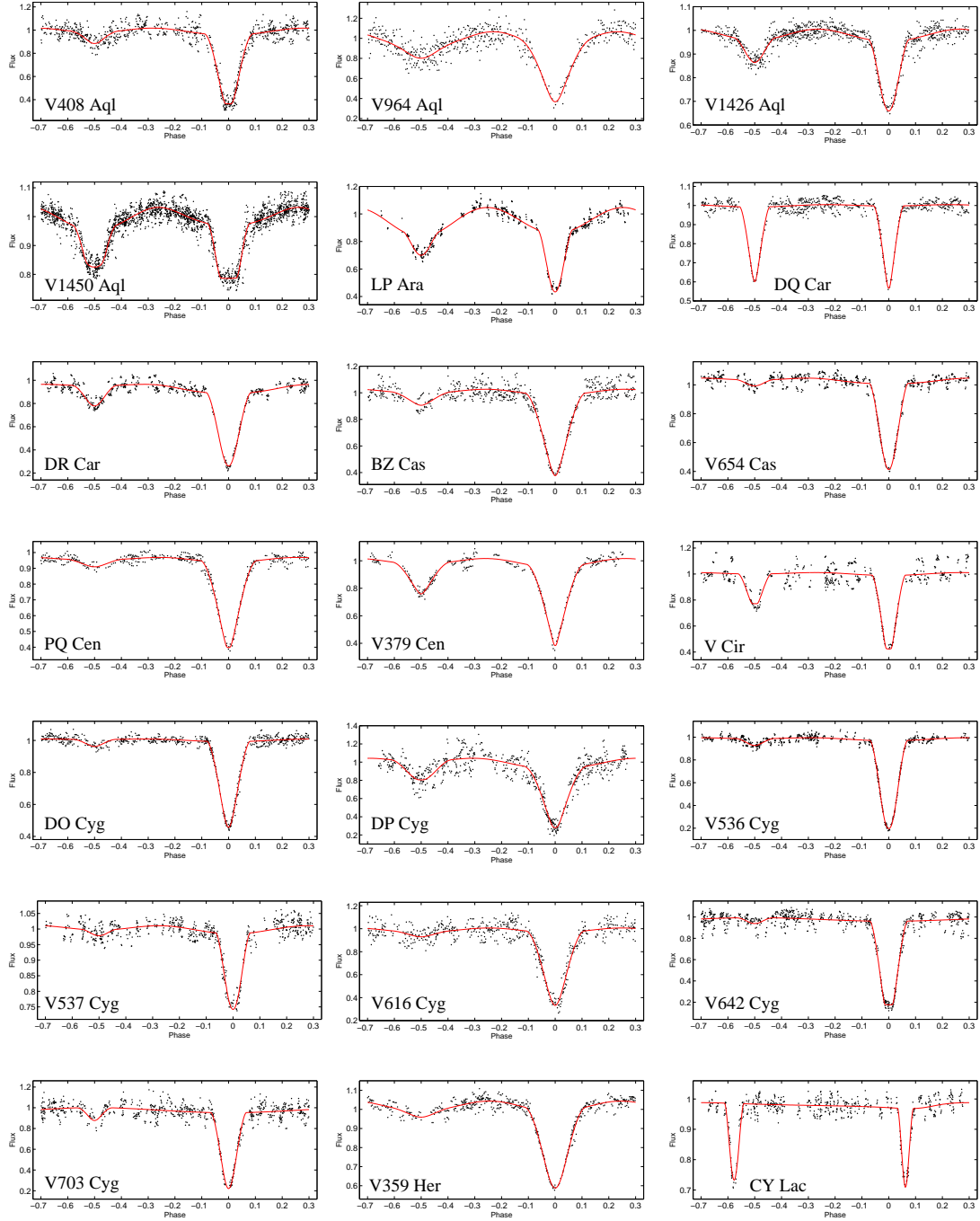}
 \caption{The light curves of the analyzed systems.}
 \label{Figs}
\end{figure}

\begin{figure}[b]
 \includegraphics[width=16cm]{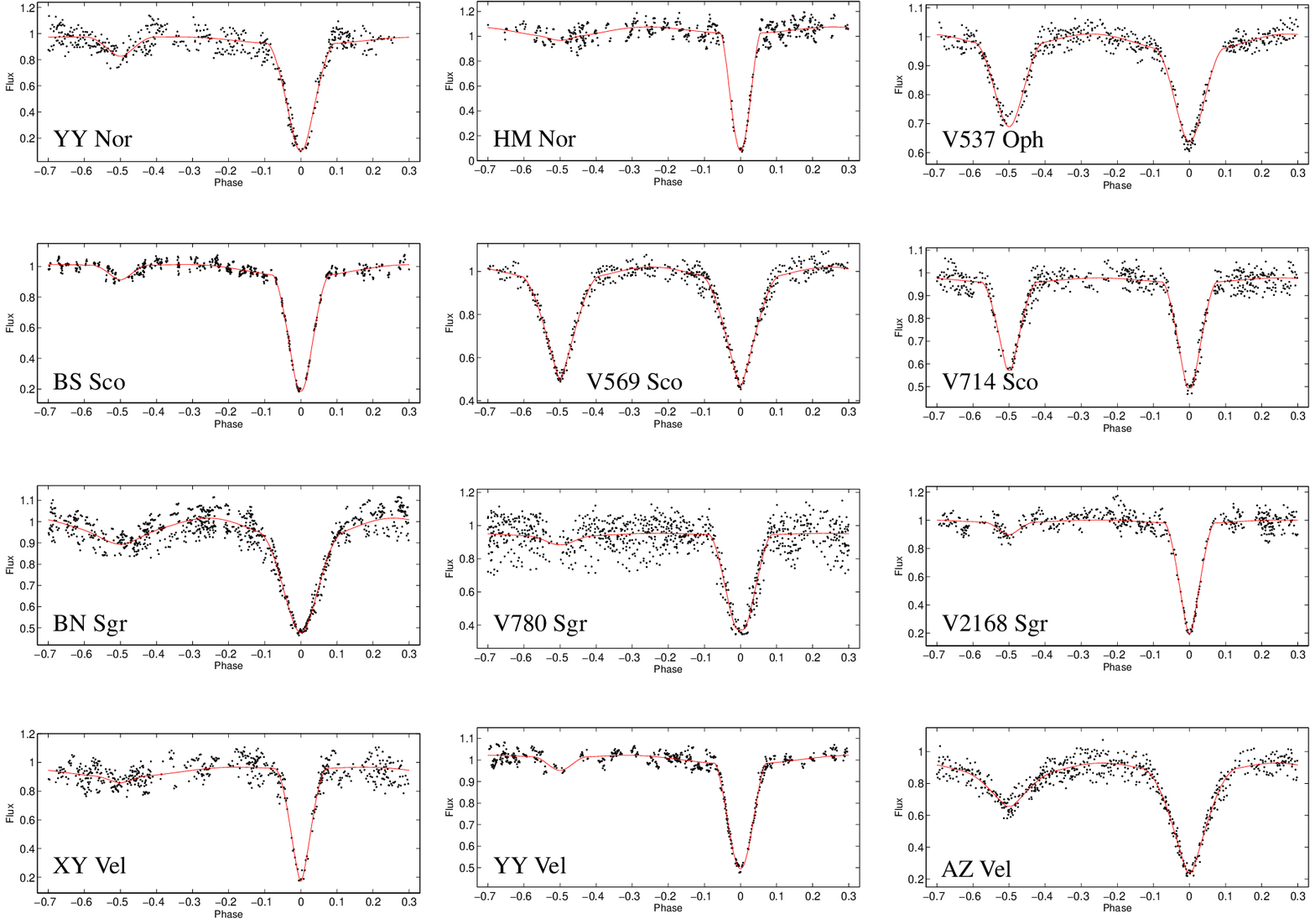}
 \caption{The light curves of the analyzed systems.}
 \label{Figs2}
\end{figure}

The results are introduced in Figs.\ref{Figs} and \ref{Figs2} and Table \ref{Table2}, where are
given all relevant parameters of the analyzed systems: HJD$_0$ and $P$ are the ephemerides of the
system, $i$ stands for the inclination, $q$ denoted the mass ratio, the 'Type' refers the mode used
for the best solution ('D' for a detached na 'SD' for a semi-detached one, see above), $\Omega_i$
stands for the Kopal's modified potentials, $T_i$ for the effective temperatures, $L_i$ for the
luminosities, $R_i$ for the radii, $F_i$ for the synchronicity parameters, and $x_i$ for the
limb-darkening coefficients (the linear cosine law was used), respectively. Inclinations smaller
than $90^\circ$ mean that the binary rotates counter-clockwise as projected onto a plane of sky.
Only two systems (V780~Sgr and AZ Vel) have their respective orbital periods shorter than 1~day and
CY~Lac was found to be the eccentric eclipsing binary. In some systems their orbital periods were
found to be different from the values published in the literature (e.g. in GCVS). The most reliable
information about its orbital elements was found in the online 'O-C gateway'\footnote{see
$\mathrm{http://var.astro.cz/ocgate/}$} \citep{Paschke}.

The parameters of CY~Lac are the following: the eccentricity $e = 0.2565$ and the argument of
periastron $\omega = 2.575$~rad. In this system both primary and secondary minima have
approximately equal depths, so the primary and secondary components (and also both minima) could be
interchanged. This is the only case, where (due to its eccentricity) the value of $HJD_0$ in Table
\ref{Table2} does not refer to the time of minimum light suitable for future observations. One time
of minimum light for this system has been derived: $2454248.8262 \pm 0.0049$.

\begin{landscape}
\begin{table}
 \scriptsize
 \caption{Basic information about the analyzed systems, taken from the literature.}
 \label{Table1} \centering \scalebox{1}{
\begin{tabular}{ c c c c c c c c c c c c c r }
\hline \hline
    Star   & Mag B & Mag V & (B-V) &  (B-V) &   Sp.  &    Sp.     &  q   &  Type  &Minima&  Mag  &  Mag  & Mag   &\multicolumn{1}{c}{Data} \\
           & GCVS  & GCVS  & GCVS  &  Nomad &        &   S\&K     & S\&K &  S\&K  &      &  OMC  &  MinI & MinII &        \\
    \hline
  V408 Aql & 14.10 &       &       &  0.790 &        &(A2)+[G8IV] & 0.170&  EA/SD &  16  & 13.08 & 14.20 & 13.23 &   822  \\
  V964 Aql & 13.20 &       &       &  0.360 &        &(A3)+[G0IV] & 0.410&  EA/SD &   4  & 13.73 & 14.89 & 14.04 &   511  \\
 V1426 Aql &  9.63 &  9.18 & 0.45  &  0.433 &   G0   &            &      &        &  10  &  9.14 &  9.60 &  9.31 &   823  \\
 V1450 Aql &  9.28 &  8.99 & 0.29  &  0.272 & A0V+A? &            &      &        &      &  8.87 &  9.16 &  9.11 &  2234  \\
    LP Ara & 10.48 & 10.2  & 0.28  &  0.253 &   B8   &  B8+[A8]   & 0.090&  EA/DS &   0  & 10.05 & 11.01 & 10.49 &   541  \\
    DQ Car & 11.3  & 11.1  & 0.2   &  0.203 &   A0   & A0+[G1IV]  & 0.240&  EA/SD &   3  & 11.04 & 11.67 & 11.60 &   567  \\
    DR Car & 11.80 &       &       &  0.060 &   B5   & B5+[G0IV]  & 0.260&  EA/SD &   0  & 11.34 & 12.76 & 11.57 &   556  \\
    BZ Cas &       & 11.4  &       &  0.399 &   A0   & A0+[G1.5IV]& 0.320&  EA/SD &  49  & 11.27 & 12.36 & 11.41 &   560  \\
  V654 Cas & 11.4  & 10.9  & 0.5   &  0.363 & B3-B5V &            &      &        &      & 10.85 & 11.86 & 10.91 &   585  \\
    PQ Cen & 10.50 &       &       &  0.192 &        & (A8)+[K0IV]& 0.260&  E/SD  &   3  & 10.63 & 11.60 & 10.70 &   379  \\
  V379 Cen &  8.80 &  8.1  & 0.70  & -0.007 &  B5Vn  &  B5V+[A3]  & 0.350&  EA/SD &   4  &  8.78 &  9.84 &  9.10 &   258  \\
     V Cir & 10.80 & 10.7  & 0.10  &  0.637 &        & (B0)+[B3]  & 0.490&  EB/DM &  26  & 10.74 & 11.68 & 11.03 &   441  \\
    DO Cyg & 11.2  & 10.7  & 0.5   &  0.346 &   A0   &  A0+[G2IV] & 0.200&  EA/SD &  83  & 10.74 & 11.60 & 10.79 &   636  \\
    DP Cyg & 13.20 &       &       &  0.130 &        & (A2)+[K6IV]& 0.200&  EA/SD &   2  & 12.85 & 14.28 & 13.13 &   641  \\
  V536 Cyg & 11.90 &       &       &  0.386 &        & (A2)+[K1IV]& 0.120&  EA/SD &  16  & 11.26 & 13.02 & 11.33 &   624  \\
  V537 Cyg & 11.4  & 10.6  & 0.8   &  0.884 &  A-B   & A(5)+[G8IV]& 0.130&  EA/SD &  20  & 10.64 & 10.97 & 10.67 &   625  \\
  V616 Cyg & 13.80 &       &       &  0.250 &        & (B7)+[F4]  & 0.370&  EA/SD &  33  & 12.89 & 14.10 & 12.98 &   544  \\
  V642 Cyg & 12.90 &       &       &  0.230 &        & (A5)+[K0IV]& 0.150&  EA/SD &  15  & 12.61 & 14.48 & 12.67 &   728  \\
  V703 Cyg & 13.50 &       &       & -0.570 &        &            &      &        &   3  & 12.90 & 14.54 & 13.04 &   692  \\
  V359 Her & 10.30 & 10.03 & 0.27  &  0.305 &   F0   &  F0+[G9IV] & 0.320&  EA/SD & 141  &  9.95 & 10.58 & 10.04 &   505  \\
    CY Lac & 11.53 & 11.32 & 0.21  &  0.041 &   B5   &  B5V+[F0]  & 0.300&  EA/SD &   0  & 11.35 & 11.71 & 11.68 &   557  \\
    YY Nor & 13.20 &       &       &  0.270 &        & (B9)+[G0IV]& 0.400&  EA/SD &   0  & 12.73 & 15.13 & 12.91 &   494  \\
    HM Nor & 11.5  & 11.1  & 0.4   &  0.400 &        & (A5)+[K0IV]& 0.150&  EA/SD &   0  & 11.22 & 14.02 & 11.34 &   496  \\
  V537 Oph & 12.50 &       &       &  0.140 &        & (F0)+[F0]  & 1.000&  EA/DW &   2  & 12.19 & 12.69 & 12.60 &   498  \\
    BS Sco & 11.1  & 10.7  & 0.4   &  0.391 &  B5V   & B5V+[G2IV] & 0.280&  EA/SD &   0  & 10.59 & 12.42 & 10.70 &   549  \\
  V569 Sco & 10.70 &       &       &  0.641 &   A3   &  A3+[A4]   & 0.940&  EA/DM &  14  & 10.68 & 11.52 & 11.46 &   520  \\
  V714 Sco & 12.20 &       &       &        &        & (A7)+[G7IV]& 0.330&  EA/SD &   0  & 11.92 & 12.66 & 12.51 &   576  \\
    BN Sgr &  9.60 &  9.28 & 0.32  &  0.566 &  G2-5  &  F6+[K0IV] & 0.400&  EA/SD &   3  &  9.28 & 10.10 &  9.42 &   869  \\
  V780 Sgr & 12.80 &       &       &  0.390 &        & (A5)+[G2IV]& 0.110&  EA/SD &   0  & 13.35 & 14.37 & 13.43 &   912  \\
 V2168 Sgr & 12.50 &       &       &  0.680 &        &            &      &        &   1  & 13.30 & 15.08 & 13.42 &   511  \\
    XY Vel & 11.50 &       &       &  0.549 &        & (A5)+[G8IV]& 0.210&  EA/SD &   4  & 11.94 & 13.81 & 12.07 &   583  \\
    YY Vel & 11.3  & 11.1  & 0.2   &  0.113 &        & (A2)+[K1IV]& 0.220&  EA/SD &   0  & 11.03 & 11.81 & 11.10 &   541  \\
    AZ Vel & 12.70 &       &       &  0.260 &        & (A8)+[F5]  & 0.500&  EA/SD &   2  & 12.58 & 14.04 & 12.96 &   620  \\ \hline
\end{tabular}}
\end{table}
\end{landscape}

\begin{landscape}
\begin{table}
 \scriptsize
 \caption{The light-curve parameters of the individual systems, as derived from our analysis.}
 \label{Table2} \centering \scalebox{1}{
\begin{tabular}{ c c c c c c c c c r r r c c c c c }
\hline \hline
 Parameter &  HJD$_0$  &      P     &   $i$  &    $q$     & Type & $\Omega_1$ & $\Omega_2$ & $T_1/T_2$ &\multicolumn{1}{c}{$L_1$}&\multicolumn{1}{c}{$L_2$}&\multicolumn{1}{c}{$L_3$}& $R_1/R_2$ & $F_1$ & $F_2$ & $x_1$ & $x_2$ \\
    Star   & 2450000+  &   [days]   &  [deg] & $=M_2/M_1$ &      &            &            &           &\multicolumn{1}{c}{[\%]}&\multicolumn{1}{c}{[\%]}&\multicolumn{1}{c}{[\%]}&           &       &       &       &       \\
    \hline
  V408 Aql & 2742.574  & 2.83503997 & 89.094 &    0.6     &  SD  &    6.0478  &     --     &   1.815   & 63.75 & 14.96 & 21.29 &   0.713   & 5.106 & 1.883 & 0.499 & 0.787 \\
  V964 Aql & 2729.309  & 1.26290000 & 85.342 &    0.5     &  SD  &    3.1582  &     --     &   0.993   & 84.54 & 11.90 &  3.57 &   1.196   & 1.485 & 0.000 & 0.690 & 0.682 \\
 V1426 Aql & 2709.109  & 1.17515945 & 77.683 &    0.7     &  SD  &    4.9504  &     --     &   1.367   & 68.58 & 29.21 &  2.21 &   0.940   & 0.000 & 3.187 & 0.472 & 0.635 \\
 V1450 Aql & 2740.317  & 4.81261051 & 93.106 &    1.5     &  SD  &   10.9790  &     --     &   1.040   & 19.41 & 73.20 &  7.39 &   0.396   & 9.906 & 2.634 & 0.457 & 0.478 \\
    LP Ara & 2674.116  & 8.53282038 & 77.079 &    0.2     &  SD  &    4.7800  &     --     &   1.143   & 70.52 & 29.48 &  0.00 &   1.135   & 3.748 & 2.316 & 0.500 & 0.500 \\
    DQ Car & 2825.284  & 1.73367847 & 92.179 &    0.8     &   D  &    7.1553  &    5.9273  &   0.841   & 52.68 & 43.71 &  3.61 &   1.036   & 5.332 & 0.000 & 0.350 & 0.309 \\
    DR Car & 3146.403  & 3.99577477 & 97.608 &    0.7     &  SD  &    5.4403  &     --     &   2.197   & 70.65 & 29.35 &  0.00 &   0.725   & 0.000 & 2.220 & 0.292 & 0.472 \\
    BZ Cas & 3551.758  & 2.12646842 & 95.307 &    0.6     &   D  &    3.7192  &    3.0535  &   1.548   & 74.75 &  9.74 & 15.51 &   0.988   & 0.000 & 0.207 & 0.319 & 0.452 \\
  V654 Cas & 2656.527  & 4.94207240 & 79.681 &    0.3     &  SD  &    5.9987  &     --     &   2.532   & 72.26 & 10.51 & 17.24 &   0.593   & 0.000 & 0.388 & 0.285 & 0.537 \\
    PQ Cen & 2831.836  & 1.05718895 & 87.357 &    0.4     &   D  &    3.4249  &    2.7701  &   2.307   & 81.34 &  3.54 & 15.12 &   1.204   & 0.000 & 0.544 & 0.500 & 0.255 \\
  V379 Cen & 2651.290  & 1.87469639 & 89.393 &    0.6     &   D  &    3.9956  &    3.2835  &   1.331   & 66.22 & 20.52 & 13.26 &   1.061   & 1.838 & 0.880 & 0.514 & 0.563 \\
     V Cir & 3062.298  & 4.40923643 & 91.678 &    0.6     &   D  &    5.5716  &    5.1156  &   1.655   & 84.16 & 15.84 &  0.00 &   1.425   & 3.755 & 0.000 & 0.074 & 0.279 \\
    DO Cyg & 4087.345  & 1.70999742 & 84.467 &    0.4     &   D  &    4.7945  &    2.9291  &   2.078   & 66.11 &  5.09 & 28.80 &   0.834   & 0.504 & 1.498 & 0.291 & 0.533 \\
    DP Cyg & 2746.748  & 2.34691815 & 86.792 &    0.3     &  SD  &    3.1773  &     --     &   1.795   & 87.17 & 12.83 &  0.00 &   1.205   & 1.272 & 0.000 & 0.529 & 0.814 \\
  V536 Cyg & 2836.598  & 6.01045459 & 84.108 &    0.5     &   D  &    6.3426  &    3.2831  &   1.727   & 82.58 & 12.81 &  4.61 &   0.656   & 1.161 & 1.826 & 0.579 & 0.500 \\
  V537 Cyg & 3347.884  & 4.75843337 & 77.032 &    0.9     &  SD  &    6.6202  &     --     &   1.863   & 65.70 &  8.52 & 25.78 &   0.766   & 4.261 & 3.605 & 0.436 & 0.707 \\
  V616 Cyg & 2653.970  & 1.32665076 & 82.405 &    0.6     &   D  &    3.9833  &    3.0757  &   1.423   & 82.80 &  5.16 & 12.04 &   0.918   & 1.111 & 0.768 & 0.502 & 0.746 \\
  V642 Cyg & 2761.959  & 4.44652373 & 90.046 &    0.8     &   D  &    6.8413  &    4.3267  &   2.549   & 81.49 &  6.35 & 12.16 &   0.656   & 2.180 & 1.552 & 0.316 & 0.611 \\
  V703 Cyg & 3125.944  & 4.14529239 & 85.315 &    0.7     &   D  &    6.4209  &    4.1021  &   2.043   & 79.22 & 16.04 &  4.74 &   0.738   & 0.000 & 0.000 & 0.440 & 0.797 \\
  V359 Her & 3574.180  & 1.75576649 & 79.273 &    0.8     &  SD  &    4.0204  &     --     &   1.350   & 65.38 &  5.34 & 29.28 &   0.955   & 1.089 & 1.687 & 0.532 & 0.757 \\
    CY Lac & 4094.814  & 8.35974636 & 83.474 &    0.6     &   D  &    8.4176  &    7.7674  &   1.498   & 57.81 & 42.19 &  0.00 &   0.672   & 3.154 & 4.941 & 0.406 & 0.549 \\
    YY Nor & 2726.466  & 1.69498989 & 87.133 &    0.6     &   D  &    4.7717  &    3.2341  &   2.097   & 83.80 & 16.20 &  0.00 &   0.774   & 0.000 & 1.271 & 0.319 & 0.522 \\
    HM Nor & 2726.068  & 4.42628455 & 76.341 &    0.2     &  SD  &    6.8694  &     --     &   1.336   & 89.57 & 10.43 &  0.00 &   0.569   & 3.199 & 0.133 & 0.544 & 0.764 \\
  V537 Oph & 3606.987  & 1.14718255 & 80.989 &    0.7     &   D  &    4.3997  &    3.7075  &   1.213   & 47.30 & 43.48 &  9.22 &   0.969   & 1.411 & 1.458 & 0.528 & 0.665 \\
    BS Sco & 2876.953  & 7.62241175 & 86.533 &    2.7     &  SD  &    8.1521  &     --     &   3.274   & 88.42 &  6.40 &  5.18 &   0.826   & 2.835 & 5.084 & 0.309 & 0.587 \\
  V569 Sco & 2673.110  & 1.04724351 & 88.826 &    1.2     &   D  &    4.5470  &    4.7905  &   0.981   & 50.78 & 49.22 &  0.00 &   0.974   & 0.894 & 0.760 & 0.454 & 0.444 \\
  V714 Sco & 2743.254  & 1.39644612 & 88.728 &    1.2     &   D  &    6.0192  &    5.7818  &   1.012   & 47.06 & 52.94 &  0.00 &   0.835   & 0.696 & 1.088 & 0.442 & 0.449 \\
    BN Sgr & 2751.347  & 2.51976721 & 76.280 &    1.0     &  SD  &    5.3627  &     --     &   1.532   & 65.84 & 11.54 & 22.63 &   0.674   & 3.282 & 0.662 & 0.531 & 0.689 \\
  V780 Sgr & 2746.065  & 0.86031002 & 84.031 &    0.9     &   D  &    5.7484  &    3.8418  &   1.069   & 59.96 & 12.28 & 27.76 &   0.636   & 0.000 & 0.319 & 0.357 & 0.378 \\
 V2168 Sgr & 3118.315  & 2.06886580 & 86.908 &    1.2     &   D  &    6.1918  &    6.5589  &   1.478   & 89.89 &  9.78 &  0.33 &   0.953   & 1.976 & 1.597 & 0.450 & 0.555 \\
    XY Vel & 2998.041  & 2.51019823 & 88.230 &    0.7     &   D  &    6.3452  &    4.5978  &   0.764   & 86.95 &  4.32 &  8.73 &   0.972   & 5.371 & 2.438 & 0.823 & 0.613 \\
    YY Vel & 2824.859  & 4.16420826 & 78.788 &    0.7     &   D  &    7.0504  &    3.6171  &   1.909   & 62.23 & 20.99 & 16.78 &   0.560   & 4.581 & 1.468 & 0.406 & 0.653 \\
    AZ Vel & 2805.448  & 0.77578092 & 89.973 &    0.5     &  SD  &    3.1843  &     --     &   0.987   & 78.45 & 21.55 &  0.00 &   1.097   & 0.000 & 0.168 & 0.406 & 0.403 \\ \hline
 \hline
\end{tabular}}
\end{table}
\end{landscape}

Another interesting fact of this sample is that about one half of the investigated systems have the
luminosity of the third unseen body above a statistically significant value about 5\%. This result
is not surprising, because e.g. \cite{Pribulla} also discovered that more than 50\% of binaries
exist in multiple systems. One could speculate about a prospective future discovery of such
components in these systems. Due to missing detailed analysis (spectroscopic, interferometric,
etc.), the only possible way how to discover these bodies nowadays is the period analysis of their
times of minima variations. In the system BZ~Cas such an analysis exists and the third body was
discovered with orbital period about 61~yr, see \cite{2007NewA...12..613E}.

\section{Discussion and conclusions}

The light-curve analyses of thirty-three selected systems have been carried out. Using the light
curves observed by the Optical Monitoring Camera onboard the INTEGRAL satellite, one can estimate
the basic physical parameters of these systems. Despite this fact, the parameters are still only
the preliminary ones, affected by relatively large errors and some of the relevant parameters were
fixed at their suggested values. The detailed analysis is still needed, especially spectroscopic
one, or another more detailed light curve one in different filters. Together with a prospective
radial-velocity study, the final picture of these systems could be done. Particularly, the systems
V1450~Aql and CY~Lac seem to be the most interesting ones. The first one is massive semi-detached
system, which shows total eclipses and the second one due to its eccentric orbit.

\section{Acknowledgments}
Based on data from the OMC Archive at LAEFF, pre-processed by ISDC. This investigation was
supported by the Research Program MSMT 0021620860 of the Ministry of Education of Czech Republic
and also by the Mexican grant PAPIIT IN113308. This research has made use of the SIMBAD database,
operated at CDS, Strasbourg, France, and of NASA's Astrophysics Data System Bibliographic Services.

\end{document}